\documentclass[a4paper,11pt]{article}
\usepackage{pos}
\usepackage{aas_macros}
\usepackage{siunitx}
\usepackage[UKenglish]{babel}
\usepackage{subcaption}

\DeclareSIUnit\year{yr}
\DeclareSIUnit\parsec{pc}

\title{Stochastic modelling of cosmic ray sources for diffuse high-energy gamma-rays and neutrinos}
 \ShortTitle{Stochastic modelling of cosmic ray sources for diffuse emissions}

\author*[a]{Anton Stall}
\author[a]{Leonard Kaiser}
\author[a]{Philipp Mertsch}

\affiliation[a]{Institute for Theoretical Particle Physics and Cosmology (TTK), RWTH Aachen University, 52056 Aachen, Germany}

\emailAdd{stall@physik.rwth-aachen.de}
\emailAdd{leonard.kaiser1@rwth-aachen.de}
\emailAdd{pmertsch@physik.rwth-aachen.de}

\abstract{Cosmic rays of energies up to a few \si{\peta\electronvolt} are believed to be of galactic origin, yet individual sources have still not been firmly identified. Due to inelastic collisions with the interstellar gas, cosmic-ray nuclei produce a diffuse flux of high-energy gamma-rays and neutrinos. \textit{Fermi}-LAT has provided maps of galactic gamma-rays at \si{\giga\electronvolt} energies which can be produced by both hadronic and leptonic processes. Neutrinos, on the other hand, are exclusively produced by the sought-after hadronic processes, yet they can be detected above backgrounds only at hundreds of \si{\tera\electronvolt}. Oftentimes, diffuse emission maps are extrapolated from GeV to PeV energies, but the sources contributing at either energies likely differ. We have modelled the production of diffuse emission from \si{\giga\electronvolt} through \si{\peta\electronvolt} energies in a Monte Carlo approach, taking into consideration the discrete nature of sources. We can generate realisations of the diffuse sky in a matter of seconds, thus allowing for characterising correlations in direction and energy. At hundreds of \si{\tera\electronvolt}, relevant for observations with LHAASO, Tibet AS-gamma, IceCube and the upcoming SWGO, variations between different realisations are sizeable. Specifically, we show that extrapolations of diffuse emission from \si{\giga\electronvolt} to \si{\peta\electronvolt} energies must fail and apply our results on the recent experimental findings.}

\ConferenceLogo{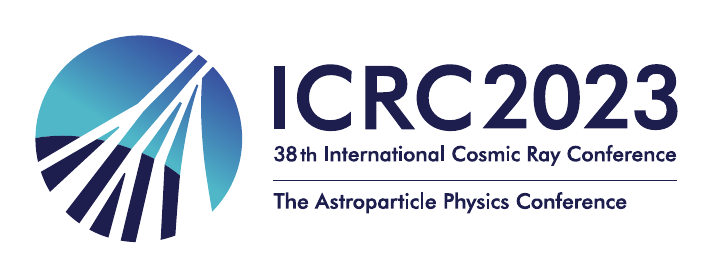}

\FullConference{%
38th International Cosmic Ray Conference (ICRC2023)\\
  26 July - 3 August, 2023\\
  Nagoya, Japan}

\begin{document}
\maketitle

\section{Introduction}
The comic ray (CR) spectrum at Earth can be measured with an ever increasing precision. Although many intriguing features have been identified \cite{Gabici2019_Origin_standardparadigm}, those measurements only capture the galactic CR flux locally. Galactic diffuse emission (GDE) is radiation produced by the interaction of CRs with the interstellar medium (ISM) \cite{Schwefer2023_CRINGE}. The observation of diffuse gamma-rays and neutrinos provides a possibility to learn more about the abundance of CRs in other regions in the Milky Way \cite{Tibaldo2021}. To predict this diffuse flux, four main ingredients are required: (1) the distribution of CR sources, (2) a suitable CR transport model, (3) detailed maps of the galactic gas, and (4) the production cross-sections of the CRs with the gas. All of those are known to be uncertain to varying degrees \cite{Schwefer2023_CRINGE}. This work focuses specifically on point (1), the source distribution. 

It is a commonly held belief that the majority of galactic CRs are accelerated through diffusive shock acceleration in supernova remnants (SNRs) \cite{Gabici2019_Origin_standardparadigm}. In that case, the spatial dimensions and activity times of the accelerators can be assumed to be much smaller than the corresponding propagation lengths and times of the CRs. Thus, SNRs are well-described by point sources that inject a burst-like spectrum at a specific point in space and time \cite{Genolini2017}. This, however, contradicts the often assumed smooth source density throughout the Galaxy in simplified models \cite{Gabici2019_Origin_standardparadigm}. It has been shown that the consideration of the point-like nature of the CR sources can lead to relevant deviations compared to a smooth source distribution, possibly giving rise to features in the CR spectrum (see e.g. \cite{Kachelriess_2015_Localsource}).

As the exact coordinates of the CR sources are not known \cite{Mertsch2011,Genolini2017, EvoliAmatoBlasi2021_Stochastic, BlasiAmato2012_1, Minh2021_forsourcerate}, the quantification of such deviations demand a probabilistic investigation of the problem. If one considers the CR spectrum at a specific position (like our solar system), one will find that the CR flux gets stronger contributions from younger and closer sources. Even more, it is possible to find realisations of the discrete source distribution that are dominated by a single local source. The important question, however, is how likely such a source realisation is. For the total local CR flux, this has been investigated theoretically and tested by Monte Carlo simulations (see e.g. \cite{Mertsch2011, Mertsch2018,Genolini2017, EvoliAmatoBlasi2021_Stochastic, BlasiAmato2012_1}). Especially, it has been pointed out that the distribution of the fluxes follows so-called stable laws with divergent variance, i.e. one has to use modified versions of the central limit theorem in a thorough theoretical treatment. In this work, we are not (only) interested in the influence of the discrete nature of sources locally. Instead, we determine the CR flux corresponding to a discrete source distribution at various positions in the Galaxy to calculate the GDE of gamma-rays and neutrinos.

We are mainly interested in the energy range above \SI{100}{\tera\electronvolt} where neutrino observatories like \textit{IceCube} are most sensitive \cite{IceCube2023_Diffuse}. The diffuse sky in gamma-rays has long been observed at \si{\giga\electronvolt} energies by Fermi-LAT \cite{Ackermann2012}. Oftentimes, the diffuse maps are extrapolated from the lower to the higher energies. However, just as it can be expected that the discrete nature of sources influences the CR flux measured locally, it is possible that GDEs are influenced by single sources. This could lead to stark deviations from the predictions of smooth source distributions and, furthermore, limit the possibility to extrapolate GDEs from \si{\giga\electronvolt} to \si{\peta\electronvolt} energies.

It is the main goal of this work to study the influence of the discrete nature of the sources. Therefore, we draw random realisations of a source distribution. Then, we calculate the CR flux in the Galaxy for various energies using an analytical solution of the CR transport equation. Finally, we calculate the GDE for the given realisation. Due to an efficient calculation of the CR fluxes and the GDEs, we can calculate this for a large number of realisations allowing us to perform statistical analyses.
\section{Method}
We built a stochastic model of CR sources for GDEs by considering protons, the main component of galactic CRs. For the choice of the galactic source, and transport parameters, we used the ones of the fiducial model in \cite{Schwefer2023_CRINGE}, which were fitted to local CR data.

The sources of galactic CRs are widely believed to be supernova remnants (SNR) \cite{Gabici2019_Origin_standardparadigm}. Their spatial dimensions and injection time-scales are small compared to the propagation lengths and times of CRs. Thus, SNRs can be considered as point sources injecting a source spectrum $Q$ at a specific position $\mathbf{x}_0$ and time $t_0$. This spectrum is assumed to only depend on the rigidity $\mathcal{R} = {p c}/{Z e}$. For simplicity, we assume that all sources lie in the galactic disk. There is a variety of radial source distribution functions for galactic CR sources $f\left(r\right)$. In this work, we assume one similar to \cite{Ferriere2001}, which has a non-vanishing source density in the galactic centre, but adapted to another Solar position at \SI{8.3}{\kilo\parsec} from the galactic centre. Furthermore, we consider a canonical source rate of \SI{0.03}{\per\year} \cite{BerghTammann1991_SNRrate,Tammann1994_SNRrate}.

For the transport of CRs in the Galaxy, we assume a simplified transport equation that considers diffusion only. This can be justified by comparing the time scales of various processes like advection or hadronic losses with the one of diffusive escape from the Galaxy. The transport equation is
\begin{equation}
    \frac{\partial \psi\left(\mathbf{x}, t, \mathcal{R}\right)}{\partial t} - \kappa\left(\mathcal{R}\right)\cdot \nabla^2 \psi \left(\mathbf{x}, t, \mathcal{R}\right) = S\left(\mathbf{x}, t\right) Q\left(\mathcal{R}\right)
\end{equation}
where $\psi\left(\mathbf{x}, t, E\right) = {d n}/{d E}$ denotes the isotropic CR density related to the differential flux $\Phi = (d^4n)/(dE \, dA \, dt \, d\Omega) = v/(4 \pi) \ \psi$ and to the phase-space density $f=\left(d^6n\right)/\left(d^3p \, d^3x\right)$ through $\psi = (4 \pi p^2)/v \ f$. Further, $S\left(\mathbf{x}, t\right)$ is the spatial and temporal distribution function for the sources and $Q\left(\mathcal{R}\right)$ is the injected source spectrum.

We solve this equation under the assumption of an isotropic diffusion coefficient $\kappa\left(\mathcal{R}\right)$ and two free escape boundary conditions at a height of $\pm z_{\text{max}}$ from the disk, i.e. $\psi\left(\pm z_{\text{max}}\right) = 0$. We do not consider a radial boundary condition. This equation can be solved analytically using the method of mirror charges. The solution for a single point source at $\left(\mathbf{x}_i, t_i\right)$, the Green's function, is given by:
\begin{equation}
    G\left(\mathbf{x}, t; \mathbf{x}_i, t_i\right) = Q\left(\mathcal{R}\right) \frac{1}{\left(2 \pi \sigma^2\right)^{3/2}} \exp\left(-\frac{\left(\mathbf{x}-\mathbf{x}_i\right)^2}{2 \sigma^2}\right) \cdot \vartheta\left(z, \sigma^2, z_{\text{max}}\right)
\end{equation}
where $\sigma^2\left(\mathcal{R}, t; t_i \right) = 2 \ \kappa\left(\mathcal{R}\right) \left(t-t_i\right)$. The first part of the equation after the source spectrum is essentially a Gaussian heat kernel that is the solution of the diffusion equation without a boundary condition. To account for the free escape boundary condition, a correction function $\vartheta$ has to be applied. This function is an infinite sum related to the Jacobi theta function \cite{Mertsch2011}. It has value in $\left[0,1\right]$ and gets much smaller than $1$ if $z$ gets close to $\pm z_{\text{max}}$ or if the diffusion length ($\approx \sigma$) is comparable with $z_{\text{max}}$.

As the discrete source distribution is a sum of $N$ point sources, the total isotropic CR density is $\psi\left(\mathbf{x}, t\right) = \sum_{i=1}^N G\left(\mathbf{x}, t; \mathbf{x}_i, t_i\right)$.
We get the CR flux $\Phi$ from the isotropic CR density by $\Phi = \frac{v}{4 \pi} \psi $. We can calculate this sum fast. On the one hand, this allows us to make Monte Carlo simulations of the CR flux for a large number of realisations. This has been done before (see e.g. \cite{Genolini2017, BlasiAmato2012_1, EvoliAmatoBlasi2021_Stochastic}). We checked whether the mean of the proton flux in the Galaxy for 100 realisations agrees with the flux for a smooth source distribution, that is obtained by evaluating a convolution integral of the source distribution function and the Green's function. The observed fluctuations decrease with the number of realisations and are below $1 \%$ for $100$ realisations. This can be regarded as a check of the method. We will not provide a statistical analysis of the distribution of fluxes here.

Instead, we want to use the CR fluxes we obtain for each realisation to calculate the corresponding GDEs. To get those, we have to perform line-of-sight integration that is in general given by
\begin{equation}\label{eq: diffuse emissions}
    J(l, b, E) = \frac{1}{4 \pi} \sum_{m,n} \int_0^{\infty} \mathrm{d} s \int_E^{\infty} \mathrm{d} E' \ \frac{\mathrm{d} \sigma_{m,n}}{\mathrm{d} E}(E', E) \ \Phi_{m}(\boldsymbol{x}, E') \ n_{\text{gas},n}(\boldsymbol{x}) \Big|_{\boldsymbol{x} = \boldsymbol{x}(l, b, s)}
\end{equation}
where $\Phi_m$ is the CR flux of species $m$ (in our case just protons), $(\mathrm{d} \sigma_{m,n} / \mathrm{d} E)(E', E)$ is the differential cross section for the production of neutrinos or gamma-rays of energy $E$ by inelastic collisions with gas of species $n$. Finally, $n_{\text{gas},n}(\boldsymbol{x})$ describes the 3D distribution of galactic gas of species $n$ (mostly atomic and molecular hydrogen) \cite{Schwefer2023_CRINGE}.

As all of those quantities are given on grids, the line-of-sight integration boils down to various matrix multiplications. With the use of linear algebra packages like \textit{NumPy} \cite{numpy2020}, these can be performed in a matter of seconds, making it feasible to calculate a large number of diffuse sky maps in very little time\footnote{This could be achieved especially due to the work of Leonard Kaiser in the context of his master's thesis.}. With those ingredients, we can look at the GDE in a stochastic way.
\begin{figure}
    \centering
    \includegraphics{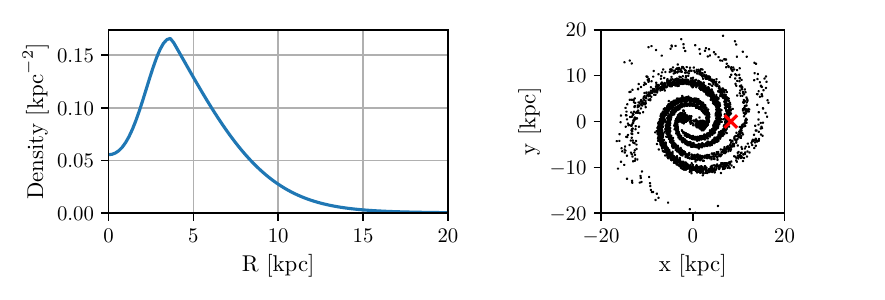}
    \caption{Radial source density similar to \cite{Ferriere2001} (left) and illustration of spiral source distribution similar to \cite{EvoliAmatoBlasi2021_Stochastic}, but with same radial source density (right). The position of the Sun is marked by a red cross.}
    \label{fig:source_distribution}
\end{figure}
\section{Results}
There are two questions we want to address in this work: (1) How do the GDEs from point-sources deviate from the ones from a smooth source distribution? and (2): How well are GDEs at \si{\giga\electronvolt} energies correlated to ones at hundreds of \si{\tera\electronvolt}? We started the investigation of these questions by analysing a set of 50 source realisations drawn from a axi-symmetric source distribution (see Figure \ref{fig:source_distribution} (left) for the radial profile). For each realisation, we determined the hadronic diffuse emissions according to equation \eqref{eq: diffuse emissions} by calculating the proton fluxes for suitable energies throughout the Galaxy. We decided to calculate the GDEs at \SI{10}{\giga\electronvolt} and \SI{100}{\tera\electronvolt}, respectively.

To answer question (1), we compared the resulting GDE sky maps with the ones for a smooth source distribution with the same radial profile and symmetry. As a check, we determined the relative deviation of the mean sky map of the 50 realisations to the sky map for the smooth source distribution. We found that at \SI{10}{\giga\electronvolt} the deviation stays below $0.7 \%$ with a standard deviation ranging from $0.5$ to $2.9 \%$ of the smooth values in different regions of the sky. For the GDEs at \SI{100}{\tera\electronvolt}, these deviations range from $-2.3 \%$ up to $1.5 \%$ with standard deviations from $2$ to $13.7 \%$ of the smooth values in different regions of the sky. We found a slight shift of $0.1$($1.7$)$\%$ in the low (high) energy case towards lower emissions that we attribute to a maximum age cut-off implemented in order to achieve finite computation times.

Although this answers question (1) to some extent, there is a caveat. The diffusion transport equation does not respect causality. To correct this flaw in the model, it was suggested \cite{Genolini2017} to implement a light-cone condition. Each source that lies outside the past light-cone of an observer will not be considered in the calculation of the flux. This will cut out some regions of the phase space and thus lead to a slightly lower proton flux. The extent of this deviation depends on the respective energy range. The higher the energy, the higher the shift. This happens as young sources make up an increasing percentage of the total flux at higher energies. Implementing the causality condition leads to a shift in the mean fluxes of between $1$ and $2\%$ for GDEs at \SI{100}{\tera\electronvolt} and is negligible ($\ll 1\%$) for GDEs at \SI{10}{\giga\electronvolt}. To account for this shift, we look at deviations from the stochastic mean rather than from the maps of the smooth (non-causal) proton fluxes.
\begin{figure}
     \centering
     \begin{subfigure}[b]{0.3\textwidth}
         \centering
         \includegraphics[width=\textwidth]{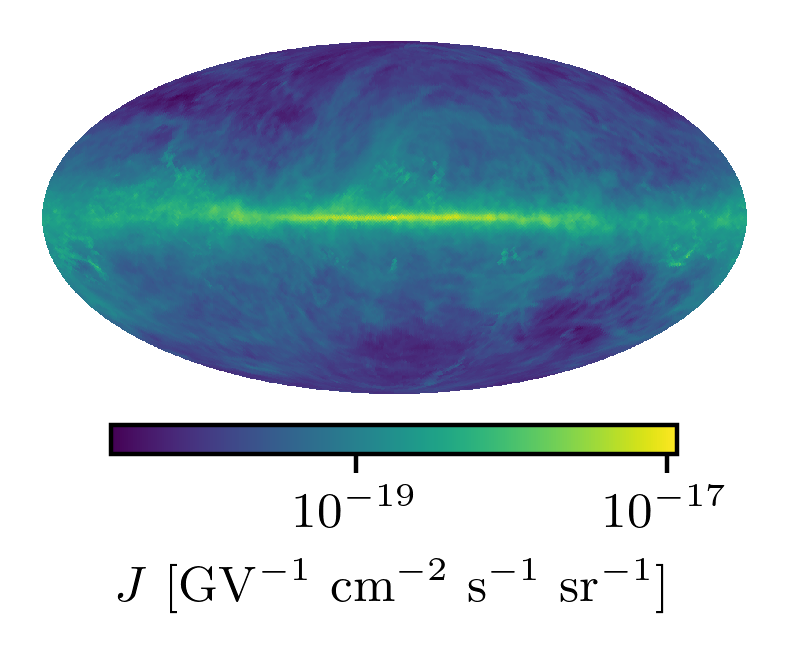}
         \caption{Sky map at \SI{100}{\tera\electronvolt}}
         \label{fig:a sky map}
     \end{subfigure}
     \hfill
     \begin{subfigure}[b]{0.3\textwidth}
         \centering
         \includegraphics[width=\textwidth]{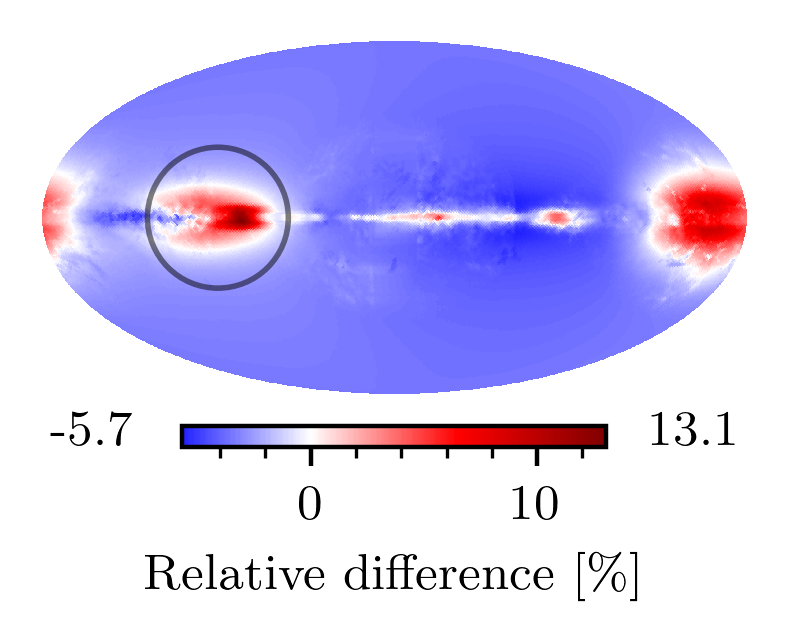}
         \caption{Relative difference to mean}
         \label{fig:b rel diff}
     \end{subfigure}
     \hfill
     \begin{subfigure}[b]{0.3\textwidth}
         \centering
         \includegraphics[width=\textwidth]{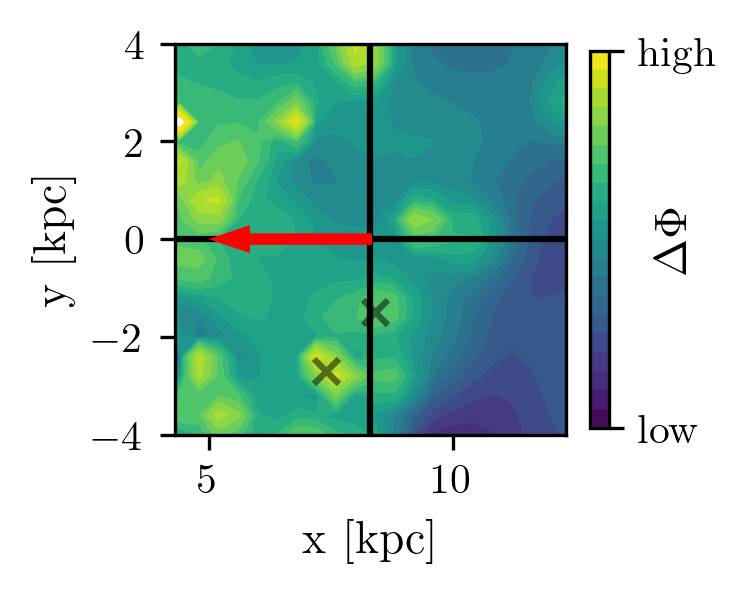}
         \caption{Deviation of local flux}
         \label{fig:c Deviations in local flux}
     \end{subfigure}
        \caption{For one specific realisation, we show the corresponding sky map of hadronic GDEs (a), the relative difference of these emissions to the stochastic mean of the emissions (b), and the projection of proton fluxes to the x-y-plane at the most contributing energy of \SI{500}{\tera\electronvolt} corrected by the smooth map to see the excess emissions in this realisation. The red arrow points towards the galactic centre. The region marked in (b) by a grey circle is likely caused by the excess proton flux marked by the crosses in (c).}
        \label{fig:one_real}
\end{figure}

In Figure \ref{fig:one_real}, we show what can be learned from such a realisation. Figure \ref{fig:a sky map} depicts a Mollweide sky map on a logarithmic scale. Differences can be studied by analysing the relative deviation from the stochastic mean of the 50 realisations. In this case, we observe a deviation of up to $13.1 \%$ in some parts of the sky which can be attributed to deviations in the proton flux around the Earth's position in the Galaxy. This can be seen by comparing Figure \ref{fig:b rel diff} with Figure \ref{fig:c Deviations in local flux}, noting that the red arrow points towards the galactic centre which is depicted in the centre of the sky map. Regions with higher GDEs can be explained by regions of increased proton flux. One prominent example of such a correspondence is marked by the grey circle and crosses. Within the 50 realisations, positive deviations of the GDEs at \SI{100}{\tera\electronvolt} got as extreme as over $30 \%$, but the median lies at $10.6 \%$. For GDEs at \SI{10}{\giga\electronvolt}, the deviations are much smaller with a median of the maximum positive deviation of $3.5 \%$.

\begin{figure}
     \centering
     \begin{subfigure}[b]{0.49\textwidth}
         \centering
         \includegraphics[width=\textwidth]{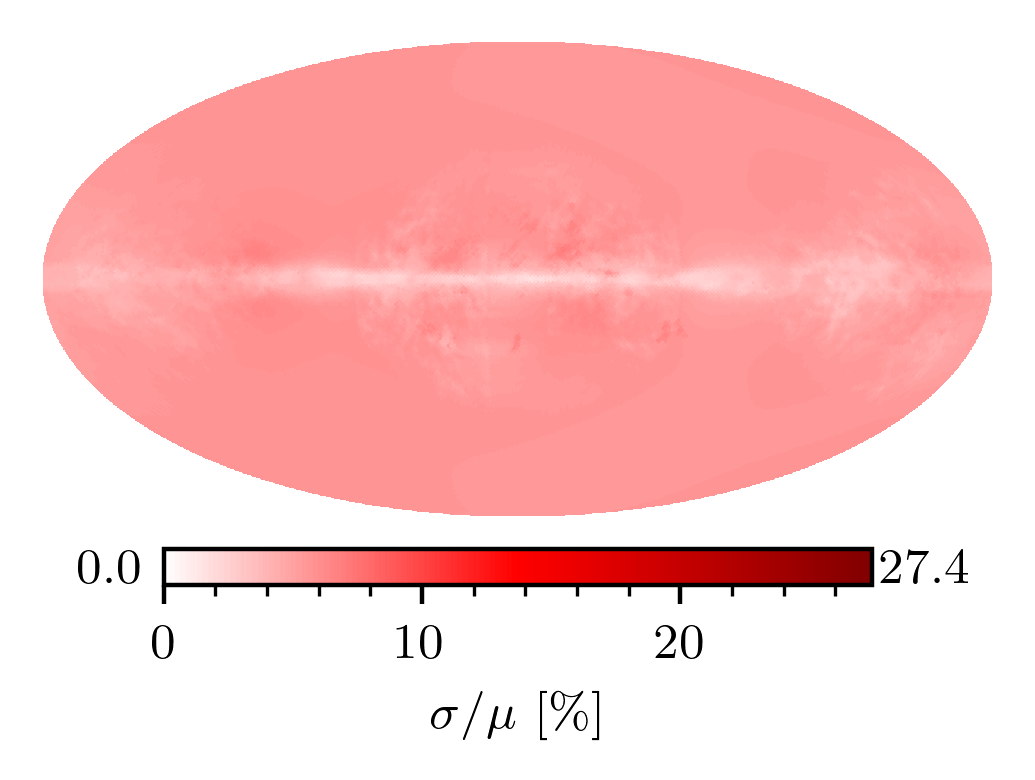}
         \caption{Axi-symmetric model}
         \label{fig:a cringe_causal}
     \end{subfigure}
     \hfill
     \begin{subfigure}[b]{0.49\textwidth}
         \centering
         \includegraphics[width=\textwidth]{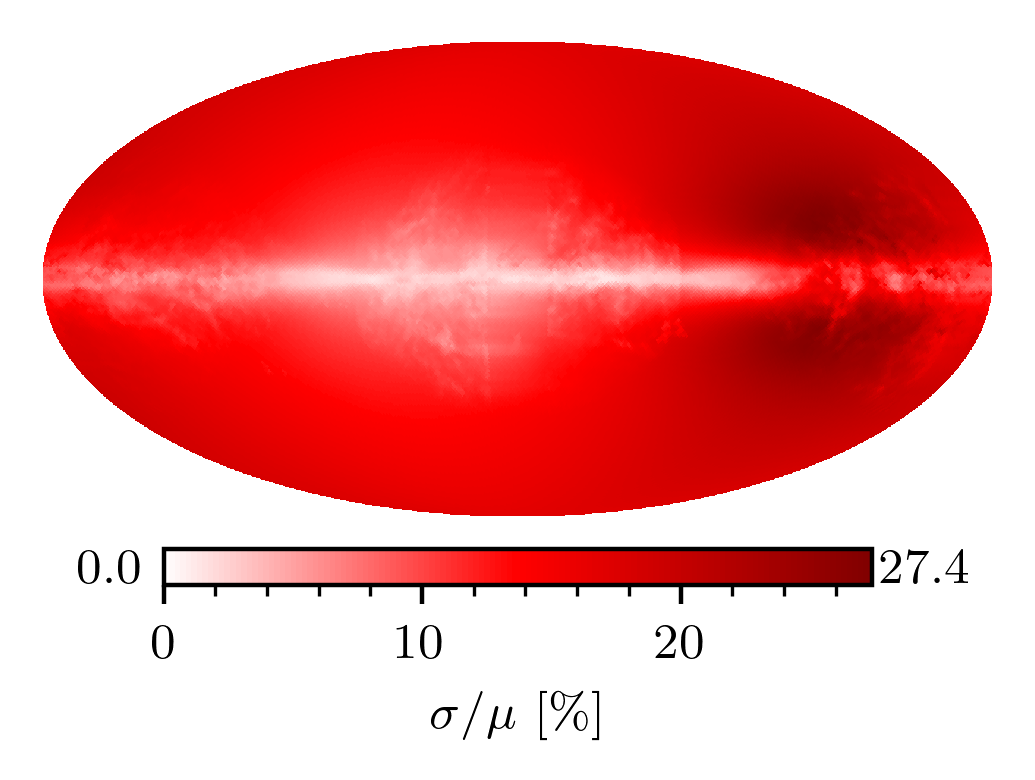}
         \caption{Spiral model}
         \label{fig:b spiral_causal}
     \end{subfigure}
        \caption{These plots depict the standard deviation of the 50 realisations normalised by their mean for different regions of the sky. In (a), the variability reaches a maximum value of $7 \%$.}
        \label{fig:spreads}
\end{figure}

The deviations of the GDEs depend strongly on the local proton flux which is mainly determined by young local sources in the  galactic neighbourhood of the solar system. Thus, it must be asked how the observed deviations depend on the more detailed distribution of sources. As has been pointed out in \cite{Mertsch2011} and \cite{EvoliAmatoBlasi2021_Stochastic}, a well-motivated specification of the SNR distribution is to assume that it follows the spiral arms of the Galaxy. We use a model for those, that is based on the one in \cite{EvoliAmatoBlasi2021_Stochastic}, but with the same radial profile as before. A depiction of this can be seen in Figure \ref{fig:source_distribution} (right). We repeated the same analysis for 50 realisations of the spiral distribution. We find that the deviations barely change for GDEs at \SI{10}{\giga\electronvolt}, and although the median for the deviations at \SI{100}{\tera\electronvolt} changes only to $10.7 \%$, larger deviations become more frequent. As can also be seen in Figure \ref{fig:spreads}, the spiral distribution increases the local variability of GDEs.

\begin{figure}
     \centering
     \begin{subfigure}[b]{0.49\textwidth}
         \centering
         \includegraphics[width=\textwidth]{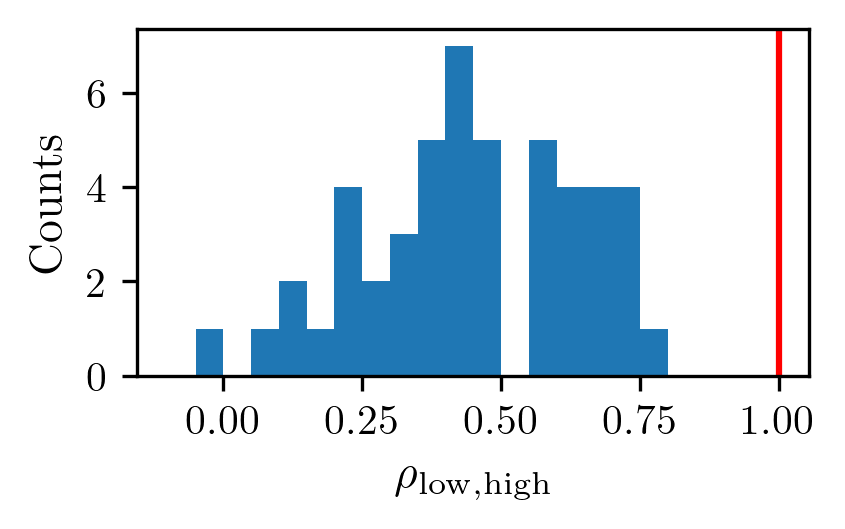}
         \caption{Axi-symmetric model}
         \label{fig:a correlation_cringe}
     \end{subfigure}
     \hfill
     \begin{subfigure}[b]{0.49\textwidth}
         \centering
         \includegraphics[width=\textwidth]{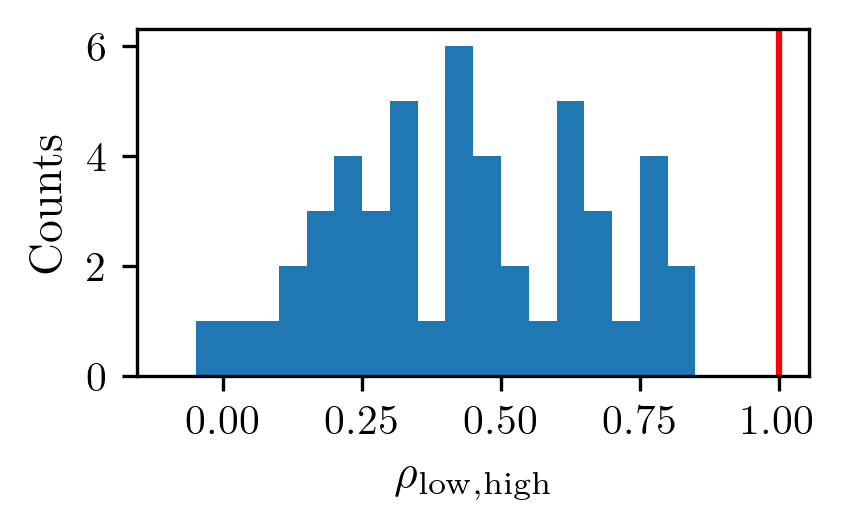}
         \caption{Spiral model}
         \label{fig:b correlation_spiral}
     \end{subfigure}
        \caption{Histogram of the correlation coefficients of the GDEs at \SI{10}{\giga\electronvolt} and \SI{100}{\tera\electronvolt} for the $50$ realisations in the axi-symmetric (a) and the spiral (b) model. The red line represents the correlation coefficient of the mean at \SI{10}{\giga\electronvolt} and \SI{100}{\tera\electronvolt} and lies almost exactly at $1$.}
        \label{fig:correlations}
\end{figure}

Now, we want to dedicate the rest of this section to question (2), the correlation of fluxes at low and high energies. Oftentimes, the diffuse maps are extrapolated from the lower to the higher energies by assuming that $J_{\text{high}} = X \cdot J_{\text{low}}$ with a fixed factor $X$. This holds true for the GDEs calculated from the smooth source distribution at \SI{10}{\giga\electronvolt} and \SI{100}{\tera\electronvolt} where this factor only varies by less then $0.1 \%$. This leads to the following equation (bar represents mean)
\begin{equation}\label{eq:correlation_assumption}
    J_{\text{low}} = \bar{J}_{\text{low}} + \Delta J \quad \Longrightarrow \quad J_{\text{high}} = X \cdot \bar{J}_{\text{low}} + X \cdot \Delta J = \bar{J}_{\text{high}} + X \cdot \Delta J
\end{equation}
Thus, we expect that $(J_{\text{low}} - \bar{J}_{\text{low}})$ and $(J_{\text{high}} - \bar{J}_{\text{high}})$ are perfectly correlated for each realisation. This, however, is not the case as can be seen in Figure \ref{fig:correlations}. We looked at the correlation coefficients defined by $\rho_{\text{X,Y}} = \sigma_{\text{X,Y}} / \sigma_{\text{X}}\sigma_{\text{Y}}$, where $\sigma_{\text{X,Y}}$ and $\sigma^2_{\text{X}}$ are the covariance of $X$ and $Y$ and the variance of $X$, respectively. We calculated them for each of the $50$ realisations in the axi-symmetric and the spiral model. Against the assumption in equation \eqref{eq:correlation_assumption}, the deviations from the mean are not always well-correlated and range from around $0$ up to approximately $0.8$. There is no qualitative difference between the axi-symmetric and the spiral model.

\section{Discussion and Conclusion}
The analysis described in the previous section leads us to the following conclusions:
\begin{enumerate}
    \item Considering point sources with burst-like injection leads to considerable deviations from the GDEs expected from a smooth source distribution,
    \item Deviations increase with energy and are higher for a spiral model due to the accumulation of sources in a certain region of the sky, and
    \item The diffuse sky at low and high energies is not well correlated.\label{point3}
\end{enumerate}
We want to emphasize that these conclusions, especially number \ref{point3}, have been obtained under the simplifying assumptions of (1) a single source population that can accelerate up to the CR knee, (2) burst-like injection, and (3) isotropic diffusion. We aim to extend the model in all of the three points mentioned above. The consideration of more than one source population with different energy ranges is expected to further increase the decoupling of diffuse emissions at low and high energies. It can also be expected that making energies at which sources inject CRs time-dependent as discussed in \cite{BlasiAmato2012_1} will increase this effect. Moving away from the completely analytical model, as would be necessary for point (3), might become computationally feasible once the calculations of the CR fluxes are suitably optimised. As the calculation of the flux from each source is perfectly parallelisable, the use of GPUs is an option that we would like to explore.

\bibliographystyle{JHEP}
\bibliography{references.bib}

\end{document}